\documentclass[useAMS,usenatbib]{mn2e}
\usepackage{natbib}

\newcommand*{\Msun}{\ensuremath{\mathrm{M_\odot}}}%

\usepackage{graphicx}
\title[Star Formation in XMMU J2235.3-2557 at $z=1.4$]{Suppression of Star Formation in the central 200 kpc of a $z=1.4$ Galaxy Cluster}

\author[Gr\"utzbauch et al.]{Ruth~Gr\"utzbauch$^{1,2}$\thanks{email: ruth@oal.ul.pt}, Amanda. E.~Bauer$^{3}$\thanks{email: abauer@aao.gov.au}, Inger J\o rgensen$^{4}$\thanks{email: Inger@gemini.edu}, Jesus Varela$^{5}$\thanks{email: jvarela@cefca.es}\\
$^{1}$ School of Physics and Astronomy, University of Nottingham, UK\\
$^{2}$ Center for Astronomy and Astrophysics, University of Lisbon, Portugal\\
$^{3}$ Australian Astronomical Observatory, PO Box 296, Epping, NSW 1710, Australia\\
$^{4}$ Gemini Observatory, 670 N. A'ohoku Pl., Hilo, HI 96720 \\
$^{5}$ Centro de Estudios de F\'isica del Cosmos de Arag\'on (CEFCA), Plaza San Juan, 1, Planta-2, E44001-Teruel, Spain\\
}

\begin{document}

\date{Accepted -- . Received --}

\maketitle

\begin{abstract}
We present the results of an extended narrow-band H$\alpha$ study of the massive galaxy cluster XMMU J2235.3-2557 at $z=1.39$. This paper represents a follow up study to our previous investigation of star-formation in the cluster centre, extending our analysis out to a projected cluster radius of 1.5~Mpc.  Using the Near InfraRed Imager and Spectrograph (NIRI) on Gemini North we obtained deep $H$ narrow-band imaging corresponding to the rest-frame wavelength of H$\alpha$ at the cluster's redshift. We identify a total of 163 potential cluster members in both pointings, excluding stars based on their near-IR colours derived from VLT/HAWK-I imaging.  Of these 163 objects 14 are spectroscopically confirmed cluster members, and 20\% are excess line-emitters.  We find no evidence of star formation activity within a radius of 200 kpc of the brightest cluster galaxy in the cluster core.  Dust-corrected star formation rates (SFR) of excess emitters outside this cluster quenching radius, $R_{_Q}~\sim~200$~kpc, are on average $<$SFR$>~= 2.7~\pm~1.0$~M$_\odot$yr$^{-1}$, but do not show evidence of increasing star-formation rates toward the extreme 1.5~Mpc radius of the cluster.  No individual cluster galaxy exceeds an SFR of $6~$M$_\odot$yr$^{-1}$.  Massive galaxies (log M$^\ast$/M$_\odot$~$>10.75$) all have low specific SFRs (SSFRs, i.e. SFR per unit stellar mass).  At fixed stellar mass, galaxies in the cluster centre have lower SSFRs than the rest of the cluster galaxies, which in turn have lower SSFRs than field galaxies at the same redshift by a factor of a few to 10. For the first time we can demonstrate through measurements of individual SFRs that already at very early epochs (at an age of the Universe of $\sim$4.5 Gyr) the suppression of star-formation is an effect of the cluster environment which persists at fixed galaxy stellar mass. 
\end{abstract}

\begin{keywords}
galaxies: evolution -- galaxies: high redshift -- galaxies: clusters
\end{keywords}

%%%%%%%%%%%%%%%%%%%%%%%%%%%%%%%%%%%%%%%%%%%%%%
%%                    INTRO                 %%
%%%%%%%%%%%%%%%%%%%%%%%%%%%%%%%%%%%%%%%%%%%%%%
\section{Introduction}\label{sec:intro}

High redshift galaxy clusters provide unique astronomical environments in which to investigate how stellar mass accumulated in galaxies when the universe was only a few billion years old.  Galaxies in high-density regions tend to be older and form stars on shorter time scales than those in lower-density regions \citep{Butcher84,Dressler97,Poggianti99,Thomas05}.  Therefore the dense regions of distant galaxy clusters are ideal environments not only to study the first massive galaxies in the universe, but also to study what happens to galaxies as they enter high density regions.

It is well established in the $z<1$ universe that star formation activity in galaxies decreases with increasing environmental density 
\citep{Jorgensen2005,Demarco2005,Tanaka2005,Marcillac2007,Patel2009,Finn2010}, such that the centres of galaxy clusters show populations of massive red galaxies with old stellar populations and no evidence of ongoing star formation.  The vast majority of star formation seen in low redshift galaxy clusters, therefore, comes from infalling galaxies before they are suffocated by the cluster environment \citep{Verdugo2008,Haines2010}.

There is some controversy as to whether the star formation - density relation reverses at $z>1$ \citep{Elbaz07,Cooper08,Hilton2010,Tran2010} or whether the period between $1<z<2$ is when environmental factors begin to dominate the shut down of star formation in the highest over-densities \citep{Strazzullo2010,Bau11a,Gru11}.  In the RDCS J1252.9-2927 cluster at $z = 1.24$, \citet{Tanaka2009} find a population of star-forming galaxies at the outskirts of the cluster, which are absent from the cluster core.  These results differ from the work of \citet{Hayashi10} who observe one of the most distant clusters known, XMMXCS J2215.9-1738 at $z=1.46$, and find no evidence for decreasing SFRs towards the cluster core.  

Most observations of high redshift clusters have not been able to determine SFRs uniquely for each cluster galaxy and have relied on broad-band colours to determine the behaviour of galaxies in distant clusters  \citep{Kurk2009,Tran2010,Hilton2010,Papovich2010}.  Evidence for high star-formation rates in the centres of $z = 1.4\sim1.6$ clusters was found by \citet{Tran2010} and \citet{Hilton2010}, whereas other authors report the presence of old stellar populations in high-$z$ cluster galaxies: \citet{Henry2010} identify a galaxy cluster at $z = 1.75$ via its X-ray emission and suggest that the stellar population of the brightest cluster galaxy has formed in a single burst at $z\sim5$. They also argue that a population of galaxies following the red sequence in the colour-magnitude diagram (i.e., old, passively evolving galaxies) might be already present in this cluster. A similar conclusion is reached by \citet{Tanaka2010} who find a possibly X-ray detected concentration of red galaxies at $z\sim1.6$, although \citet{Pierre11} find that the X-ray emission is dominated by a point source rather than an extended hot intracluster medium. The epoch around $z\sim1.5$ then seems to be critical to study the build up of the red sequence and the influence of the environment on the suppression of star-formation in galaxies in the early Universe.

We report here on the first observations of individual star-forming galaxies beyond the virial radius of the massive cluster XMMU J2235.3-2557 at $z=1.393$, which represents a follow-up study to our previous investigation of H$\alpha$ emission within the central $\sim$500 kpc of this cluster presented in \citet{Bau11a}.  The XMMU J2235.3-2557  (``XMMU2235'') cluster at $z=1.393$ is the most massive distant galaxy cluster known at these high redshifts.   It was discovered by the XMM-Newton X-ray space telescope by \citet{Mullis05}.   \citet{Rosati2009} confirm cluster membership of 34 individual galaxies in XMMU2235 based on spectroscopic data and both \citet{Jee2009} and \citet{Rosati2009} estimate the projected mass of the cluster to be around $9 \times 10^{14}\Msun$.  Near-IR observations were taken of the cluster by \citet{Lidman2008}, while rest-frame ultraviolet  properties have been investigated by \citet{Strazzullo2010}.
 
In this paper, we present narrow-band $H$ (1.57$\mu$m) observations of XMMU2235, which correspond to the wavelength region where galaxies in this cluster emit H$\alpha$.  We measure for the first time individual SFRs for 163 likely cluster galaxies in XMMU2235 out to a cluster radius of 1.5~Mpc.  In Section~$\ref{sec:data}$, we describe the near-infrared observations of the cluster, data reduction, and the calculation of star formation rates.  We discuss our results in Section~$\ref{sec:results}$, present a discussion in Section~$\ref{sec:discussion}$, and end with a summary and conclusions in Section~\ref{sec:summ}.    

Throughout the paper we assume the standard $\Lambda$CDM cosmology, a flat universe with $\Omega_\Lambda = 0.73$,  $\Omega_M = 0.27$ and a Hubble constant of $H_0 = 72 $ km s$^{-1}$ Mpc$^{-1}$.

%%%%%%%%%%%%%%%%%%%%%%%%%%%%%%%%%%%%%%%%
%%                   Observations                   %%
%%%%%%%%%%%%%%%%%%%%%%%%%%%%%%%%%%%%%%%%

\section{Data and methods}\label{sec:data}

\subsection{Observations and data reduction}\label{sec:observations}

The observations were carried out with the Near InfraRed Imager and Spectrometer \citep[NIRI,][]{Hodapp03} on Gemini North. This instrument is equipped with a $H$ narrow-band filter at a wavelength of $\lambda=1.57 \mu$m, which corresponds to the central wavelength of the H$\alpha\lambda6563\mathrm{\AA}$ emission line at the cluster redshift of $z=1.39$. This enables us to obtain rest-frame H$\alpha$ imaging and individual star-formation rates of the cluster members with a minimum amount of integration time compared to spectroscopically derived H$\alpha$ star-formation rates. The cluster area was imaged with two different pointings of adjacent fields of $\sim$2 arcmin on each side. One pointing is centred on the brightest cluster galaxy (BCG) and the second one is located in the north-east of the cluster centre, where a second concentration of spectroscopically confirmed cluster members was found in the literature \citep{Rosati2009}.

The two pointings were obtained in two different observing programs, GN-2007B-Q-79 for the central pointing and GN-2010B-Q-75 for the pointing of the cluster outskirts. The data obtained in the first program are fully described in \citet{Bau11a}. In this paper we present the data obtained in the follow-up program observed in the period from July 2010 to September 2010 and then combine data from both programs for the analysis and discussion throughout the rest of the paper.   

We obtained $H$ narrow-band observations (corresponding to H$\alpha$), as well as $H$ broad-band observations at $\lambda=1.65 \mu$m to measure the underlying continuum. The total integration time was split into single exposures of 120 seconds in the $H$ narrow-band and 30 seconds in the $H$ broad-band filter. The field of view for each single frame is 120$\times$120 arcseconds with a pixel scale of 0.1162 arcsec/pixel. Each set of 10 frames was dithered following a specific pattern to facilitate the removal of bad pixels and cosmic rays. The total number of frames observed in the narrow-band and broad-band filters were 374 and 108 frames, respectively. 

The data for the outskirts of the cluster were reduced in an identical manner to that used for the central region, as described in \citet{Bau11a}. In brief, the data reduction steps consist of the standard bias subtraction, flat-fielding, sky subtraction and co-adding of the single exposures into a combined image that is used for the source detection. The sky subtraction is the most critical step in this procedure, since the sky background in the near-infrared is bright and variable on short time-scales.  We find that the best technique to account for these effects is to co-add the science frames of five adjacent exposures into a sky frame, after subtracting all detected sources. 

Finally the reduced science frames were combined, using only frames with a good sky subtraction and without residuals of fringing, first-frame-bias pattern or sky variability. This resulted in 362 combined individual exposures in narrow-band and 102 individual exposures combined to create the broad-band image.  The final total exposure times are 51 minutes in the $H$ broad-band image of the cluster outskirts region and 724 minutes in the $H$ narrow-band filter (compared to 55 minutes and 424 minutes of the central frame in broad- and narrow-band, respectively).

The area covered by the pointing towards the outskirts of the cluster is 2.11 $\times$ 2.29 arcminutes or 4.83 arcmin$^{2}$, which is larger than the area of the central pointing of 1.6 $\times$ 1.64 arcminutes or 2.62 arcmin$^{2}$ due to a different dither pattern used. The total covered area of the XMMU~2235 cluster is then 7.45 arcmin$^{2}$. The dithering introduces image areas with different depths, mainly around the edges of the cluster outskirts field. This is taken into account in the source extraction (see Section~\ref{sec:detection}) by using the local background rms to determine the threshold for extracting a source. The physical distance from the cluster centre which we cover with our observations is $\sim 400$~kpc in all directions and up to $\sim 1.5$~Mpc towards the north-east in the pointing of the cluster outskirts. 

\begin{figure*}
\includegraphics[width=0.545\textwidth]{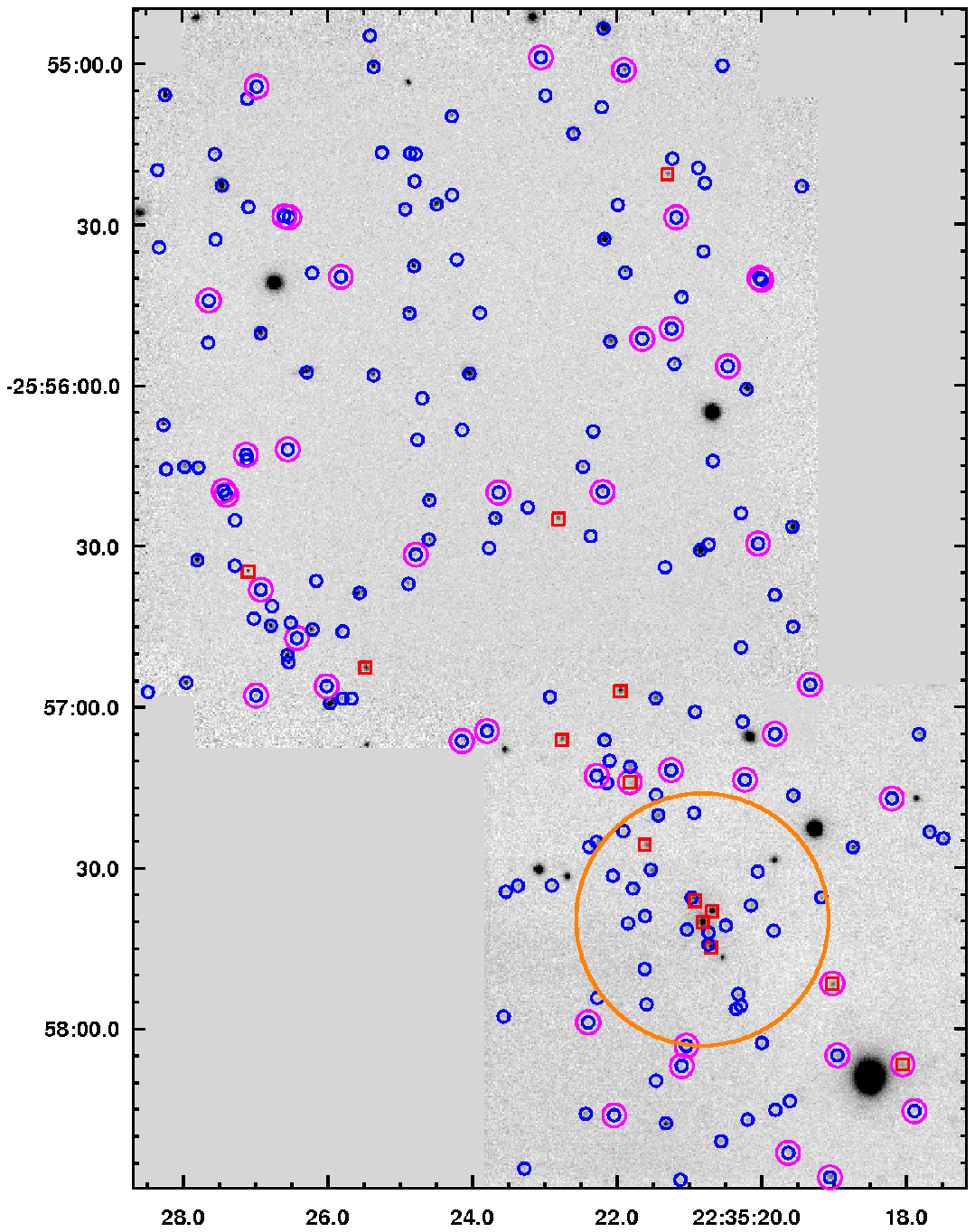}
\raisebox{0.03\height}{\includegraphics[width=0.35\textwidth]{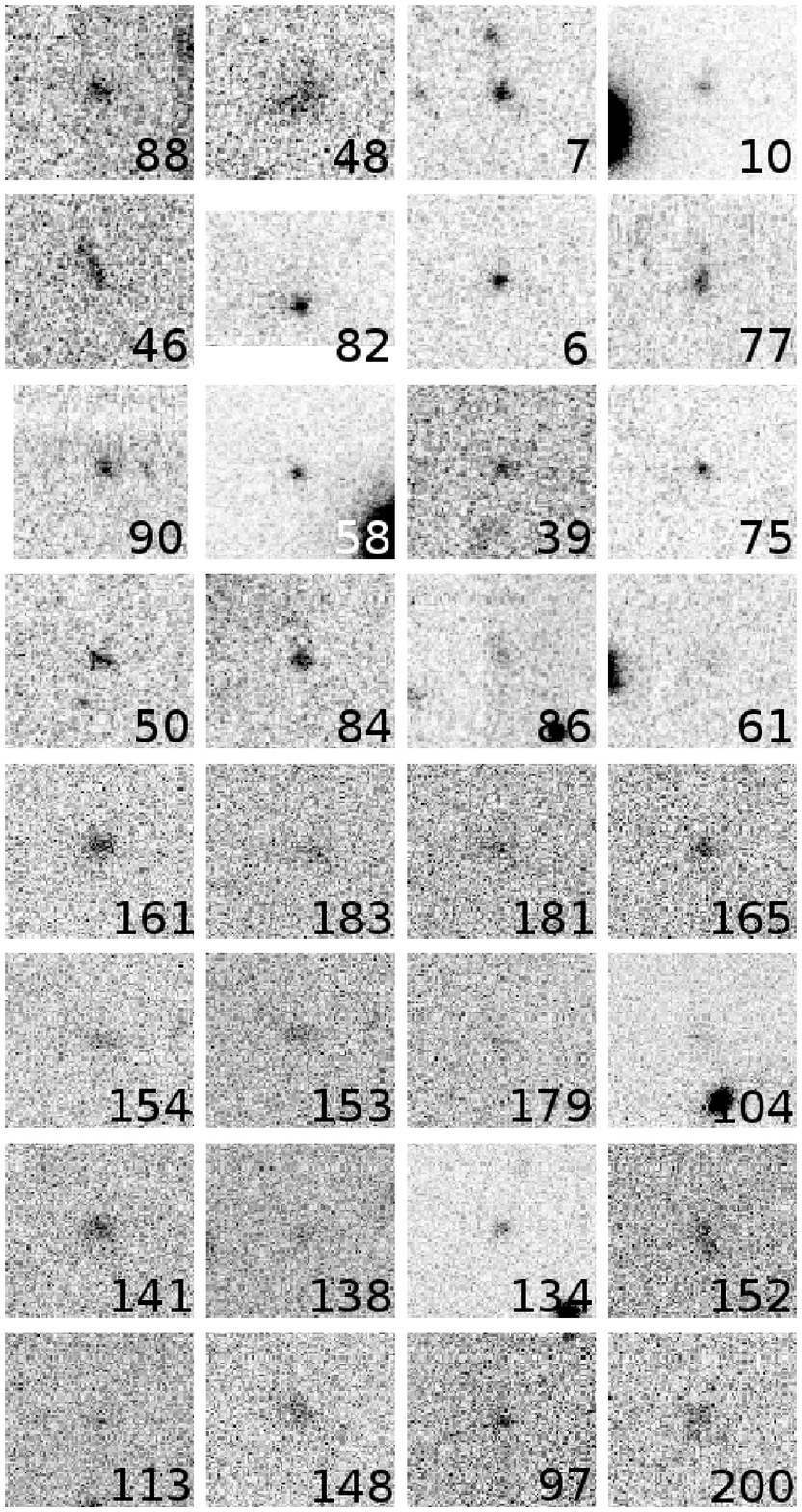}}
\caption{Final reduced, combined $H$ broad-band image of XMMU2235 (left) and $H$ narrow-band postage stamps of the 32 excess emitters (right). Left panel: Both pointings are combined in this figure. North is up, east to the left. The spectroscopically confirmed cluster members are marked with red boxes; all other detected objects are marked with blue circles. Objects additionally outlined in magenta are excess emission-line galaxies and most likely to be cluster members (see Figure~\ref{fig:ha_excess}). The orange circle indicates the radius of 200~kpc ($R_{_Q}$) within which no star-formation is found (see Section~\ref{sec:sfr-mag}. Right panel: the size of each cut-out is 10 arcseconds, corresponding to about 85 kpc on each side. The object ID (see Table~\ref{tab1}) is indicated in each image. \label{fig:images}}
\end{figure*}

Figure~\ref{fig:images} shows the $H$ broad-band image of the total covered area of both pointings. The pixel scale of 0.1162 arcsec/pixel and the angular distance scale at $z=1.39$ of 8.618 kpc/arcsec give a scale of about 1 kpc/pixel. The right panel of Figure~\ref{fig:images} shows $H$ narrow-band postage stamps of all objects identified as narrow-band excess emitter (see Section~\ref{sec:contaminants}), which are the most likely objects to be cluster members, as will be further discussed in Section~\ref{sec:contaminants}. The size of each stamp is 10~arcseconds corresponding to about 85~kpc on each side. The images indicate the presence of varied morphologies from very compact to more extended objects. However, we do not investigate the morphologies of the excess emitters here, since the resolution of our images is not high enough to reliably determine galaxy morphologies. This will be the focus of a future study using high-resolution imaging data from the Hubble Space Telescope.

%--------------------------- begin Figure 1: Images ------------------------------------------
%----------------------------- end Figure 1 ----------------------------------------

\subsection{Source detection, completeness limits and star-formation rates}\label{sec:detection}

The source detection is done separately in the two pointings due to different exposure times and noise characteristics of the two images. The photometry of galaxies in the central pointing we use here is derived by \citet{Bau11a}. To obtain the photometry of galaxies in the cluster outskirts pointing we use the same detection parameters as described in \citet{Bau11a}.
Sources are detected in both, the $H$ broad-band and narrow-band images using {\tt SExtractor} (Bertin \& Arnouts 1996). To ensure that the different depth of broad-band and narrow-band images does not affect the results, fluxes are measured within the same fixed aperture in both images. We do not apply a correction for differences in the point-spread function (PSF) since both images have a very similar PSF of $\sim 0.55$ arcseconds.
 We run {\tt SExtractor} in dual-mode, i.e. simultaneously on the broad-band and narrow-band image, detecting sources in the broad-band image and measuring photometry using the same parameters in both images. The same procedure is repeated using the narrow-band image as the detection image, in order to detect pure emission line sources that might be too faint to be detected in broad-band. The two catalogs are then combined to obtain the final source catalog shown in Table~\ref{tab1}.  The procedure of detecting sources in both bands ensures that our sample is not biased towards H$\alpha$ emitting galaxies. 

After excluding obvious spurious sources, known foreground galaxies and saturated stars, we obtain 82 objects in the central pointing (including cluster member candidates and confirmed members) and 119 objects in the outskirts pointing. We identify in total 14 spectroscopically confirmed cluster members taken from \citet{Mullis05}, \citet{Lidman2008}, and \citet{Rosati2009}, by comparing the published images (Figure~1 in each paper) to our Figure~\ref{fig:images}. Using our detection criteria (requiring 3 adjacent pixels to have a value $> 3\sigma$ of the background variation), all objects detected in the narrow-band image (72 objects) are also detected in the broad-band image. The source-extraction on the broad-band image yielded 48 additional detections, resulting in the total number of 119 objects quoted above.

In the subsequent analysis we distinguish spectroscopically confirmed members (referred to as ``confirmed members'') and all other detections by showing confirmed members in red in the plots.  The objects surrounded by an extra magenta circle in Figure~\ref{fig:images} are galaxies identified as having excess line emission and are therefore the most likely to be cluster members, as will be discussed in Section~\ref{sec:contaminants}.

The completeness limits of our broad-band and narrow-band images are estimated by placing artificial point sources in images with the same noise characteristics and object PSF as the observed images. The artificial sources are  detected in the same way as the observed sources. The counts in both, observed and artificial image are then compared to obtain the 95\% completeness limit. This procedure is fully described in \citet{Bau11a}. We obtain a 95\% completeness limit of $H_{AB} = 24.4$ in the broad-band and $H_{narrow,AB} = 23.5$ in the narrow-band, which is very similar to the depth of the central pointing of $H_{AB} = 24.2$ and $H_{narrow,AB} = 23.4$. The similar image depths despite of the longer exposure times of the outskirts pointing is due to worse observing conditions in the second observing run.

The procedure to obtain SFRs from the calibrated narrow-band magnitudes is fully outlined in \citet{Bau11a}. After calibrating the broad and narrow-band zeropoints with standard star observations, we apply a one magnitude dust correction \citep{Kennicutt83}, a statistical correction for the contribution of [N~II] emission \citep{Tresse99} and convert the resulting $H\alpha$ fluxes to star-formation rates using the \citet{Kennicutt94} relation.
The transformation between luminosity and star formation rate is known to vary by a factor of $\sim2.5$, introducing an error of $\sim30\%$.  This was accounted for in the errors given for all SFRs reported in Table~\ref{tab1}.  Note, however, that this does not include any uncertainty introduced by the extinction correction, which we cannot constrain well with the present data. Based on the magnitude limits described above in Section~\ref{sec:detection}, we find a $5~\sigma$ limiting H$\alpha$ luminosity of $L_{H\alpha} = 1\times10^{41}$~erg~s$^{-1}$, which corresponds to a $5~\sigma$ SFR$~=1.1~$M$_\odot$yr$^{-1}$.

\subsection{Cluster membership through excess emission}\label{sec:contaminants}

Our total sample consists of 201 objects detected at the 3$\sigma$ level over the total area covered in two pointings with the Gemini North telescope.  Of these, 14 are confirmed cluster members based on spectroscopic observations.  H$\alpha$ emitters at the redshift of the cluster are identified among the remaining 187 candidate cluster members as having excess flux in the narrow-band image relative to the broad-band observations.  

\begin{figure}
\includegraphics[width=0.46\textwidth]{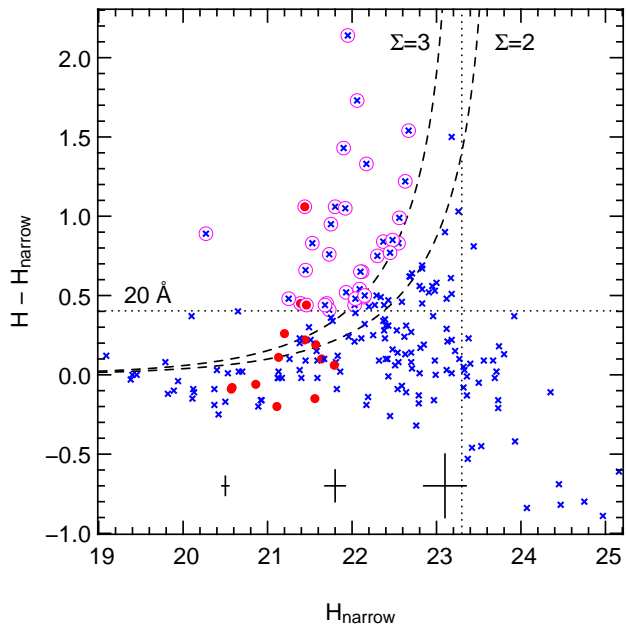}
\caption{Colour-magnitude diagram for all 201 sources detected in the narrow-band image.  The dashed lines indicate lines of constant $\Sigma$, which show the excess flux in the narrow-band image above the noise in the broad-band image.  Spectroscopically confirmed cluster member galaxies are plotted as large red points while all other detected objects are shown as blue crosses.  The vertical line shows the 95\% completeness level, so only candidates brighter than this limit are considered as possible excess emission-line objects.  Candidate emission-line cluster galaxies are objects with $EW > 20 \mathrm{\AA}$ and $\Sigma > 3$ (open magenta circles). Typical error bars are shown in bins of magnitude. \label{fig:ha_excess}}
\end{figure}

Figure~\ref{fig:ha_excess} shows the excess flux in the narrow-band filter as a function of magnitude for all 201 detected objects.  We require the narrow-band flux to be a factor of three $\Sigma$ greater than the noise in the broad-band (continuum) image \citep{Bunker1995} for an object to be considered an excess line-emitter.  The dotted curves in Figure~\ref{fig:ha_excess} show lines of constant $\Sigma$.  Emission-line candidates with low equivalent widths ($EW$) are likely to be foreground interlopers \citep{Palunas2004}, so we also apply a minimum $EW$ for the H$\alpha$ line of 20$\mathrm{\AA}$ \citep{Hayes2010}.  Candidate emission-line cluster galaxies are objects with $EW > 20 \mathrm{\AA}$ and excess line-emission of $\Sigma > 3$, which are identified as open magenta circles in Figure~\ref{fig:ha_excess}.  

Of all the detected objects, 32 satisfy the criteria to be considered excess line-emitters and are therefore more likely to be members of the cluster. Postage stamps of these 32 galaxies are shown in the right panel of Figure~\ref{fig:images}. The red points in Figure~\ref{fig:ha_excess} show spectroscopically confirmed cluster members, which are all on the bright end ($H_{narrow}<22$) of the distribution of narrow-band magnitudes of detected objects.  Even though these objects are spectroscopically confirmed as cluster members, only three of the 14 satisfy the criteria for being excess H$\alpha$ line-emitters associated with the cluster.  It is clear that while this method is a good test to identify the most likely cluster candidate members with excess line emission when no spectroscopic confirmation is available, if a galaxy in the cluster does not have excess line emission (i.e. it is not star-forming or an AGN), it will not be identified as a member of the cluster by this technique.  \citet{Lidman2008} argue that galaxies in the central region of XMMU~2235 are very likely to be cluster members based on the near-infrared colour-magnitude relation and the low likelihood of finding foreground or background galaxies in the narrow colour interval.   We will investigate the likelihood of cluster membership based on colour in the next subsection.
  
There remains potential for galaxies identified as having excess emission in H$\alpha$ at $z=1.4$, to instead be interloping objects at different redshifts, for instance they could be galaxies with [OII] emission at $z\sim3.2$ or [OIII] and H$\beta$ at $z\sim2.2$. Although a high fraction of contamination by a high redshift galaxy population seems unlikely due to their generally faint $H$ narrow-band magnitudes, there still might be a significant amount of contamination in the sample of excess emitters.  For instance \citet{Sobral11} simultaneously analysed H$\alpha$ and [O~II] narrow-band emitters in an untargeted survey at a redshift of $z=1.47$ and found the interloper fraction to be $\sim 30-40\%$. Since we are studying one of the most massive concentrations of galaxies in the early Universe, we expect the fraction of contaminants to be lower than in a general field sample. Additionally, we consider candidates that are most likely to be cluster members based on their near-infrared colours (Section~\ref{sec:redsequence}) separately in the analysis to minimize the effect of contamination on our results.

Galaxies that are significantly fainter in the narrow-band than expected from their broad-band magnitude can be seen in Figure~\ref{fig:ha_excess} as galaxies with negative $(H - H_{narrow})$ colours, which would subsequently result in negative SFRs. Note however that we have set those values to SFR=0 to avoid unphysical negative SFRs. The faint narrow-band magnitudes could be caused by stellar absorption of H$\alpha$. We have computed the $(H - H_{narrow})$ colours expected from stellar absorption using the SED of Vega and find a value of $(H - H_{narrow}) = -0.13$. This corresponds well to the measured negative colours of the brighter galaxies in our sample, for example the four bright confirmed members in the cluster centre, which show values of $(H - H_{narrow})$ from -0.06 to -0.2. Note that the expected negative colours are also in the order of the typical photometric error of $\sim0.1$ magnitudes.
The presence of H$\alpha$ absorption was also suggested by \citet{Rosati2009}, who found evidence for the presence of older stellar populations in the stacked spectra of cluster members in the cluster centre. However, the faint narrow-band magnitudes could also be caused by a strong non-H$\alpha$ absorption feature in foreground galaxies.  Galaxies at $z\sim0.85$ would have calcium triplet absorption lines at  $\lambda$$\lambda$$\lambda$8498,8542,8662$\mathrm{\AA}$ that would fall into the $\lambda 1.57 \mu$m  narrow-band filter.

%--------------------------- begin Figure 2: Halpha excess------------------------------------------
%----------------------------- end Figure 2  ----------------------------------------

\subsection{Cluster membership through colour selection}\label{sec:redsequence}

We now assess the likelihood of the cluster member candidates through their broad-band $(J-H)$ and $(H-K)$ colours  and their location in the $(J-K)$ colour-magnitude diagram. For this purpose we use additional $J$ and $K$ band imaging obtained with the HAWK-I instrument. This data is fully described in \citet{Lidman2008}. 
The source detection and photometry was carried out using the same parameters as for the $H$-band NIRI data described above, using an aperture with one arcsec radius to extract magnitudes in $J$ and $K$. The sources were then matched to the NIRI catalogue with a 1.5 arcsec matching radius. We find an average separation between matching sources of 0.25 arcsec with a rms of 0.17 arcsec. The resulting $(J-K)$ colours for all objects are given in Table~\ref{tab1}. The images do not overlap entirely with our $H$-band data and objects that are outside the field of view of the HAWK-I images are given a value of $(J-K)=99$. Only nine objects from our total narrow-band sample are not observed in $J$ and $K$, while all of the remaining source from our sample are detected in $J$ and $K$.

%--------------------------- begin Figure 3: IR colours ------------------------------------------
\begin{figure*}
\includegraphics[width=0.46\textwidth]{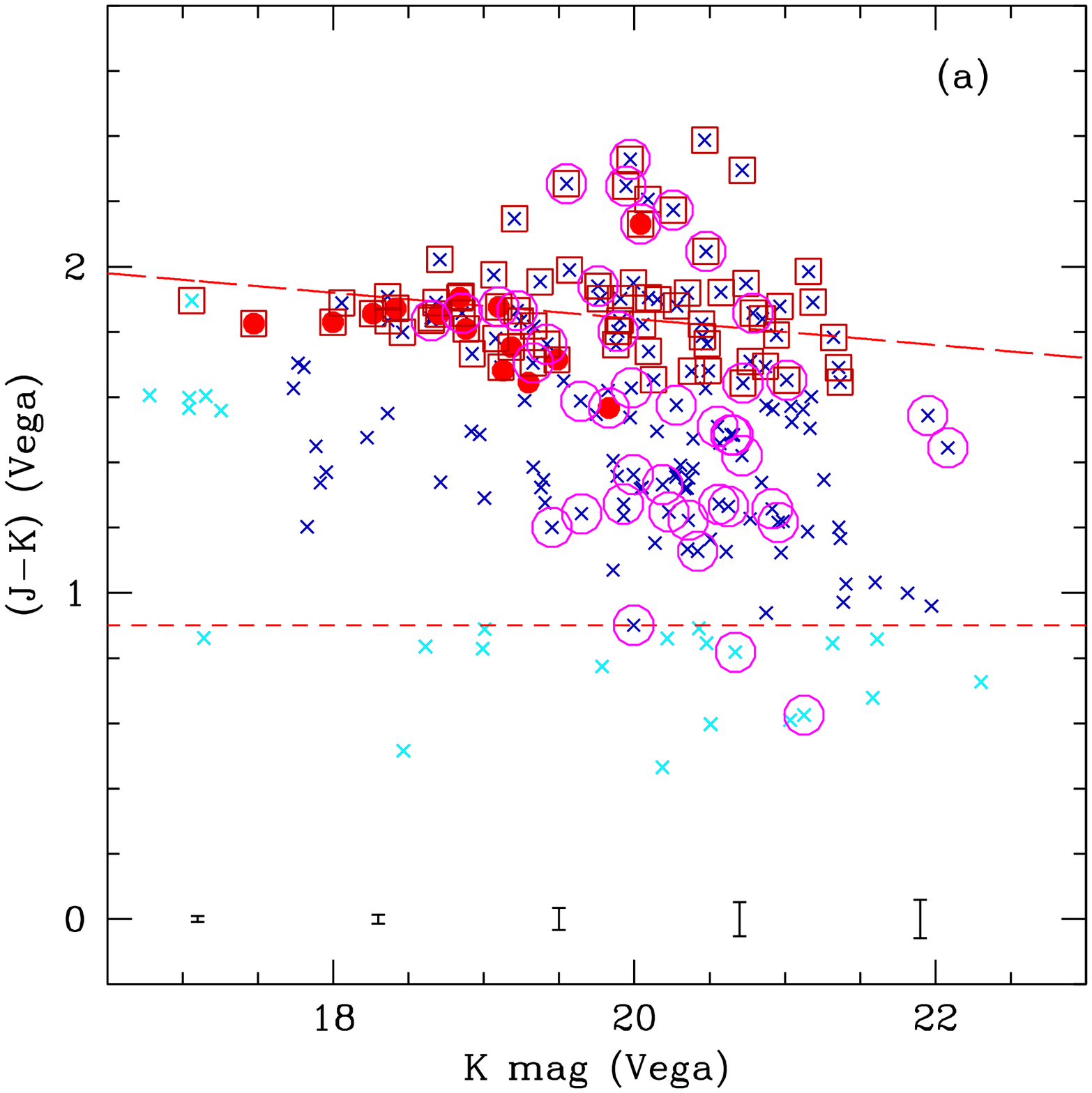}
\includegraphics[width=0.46\textwidth]{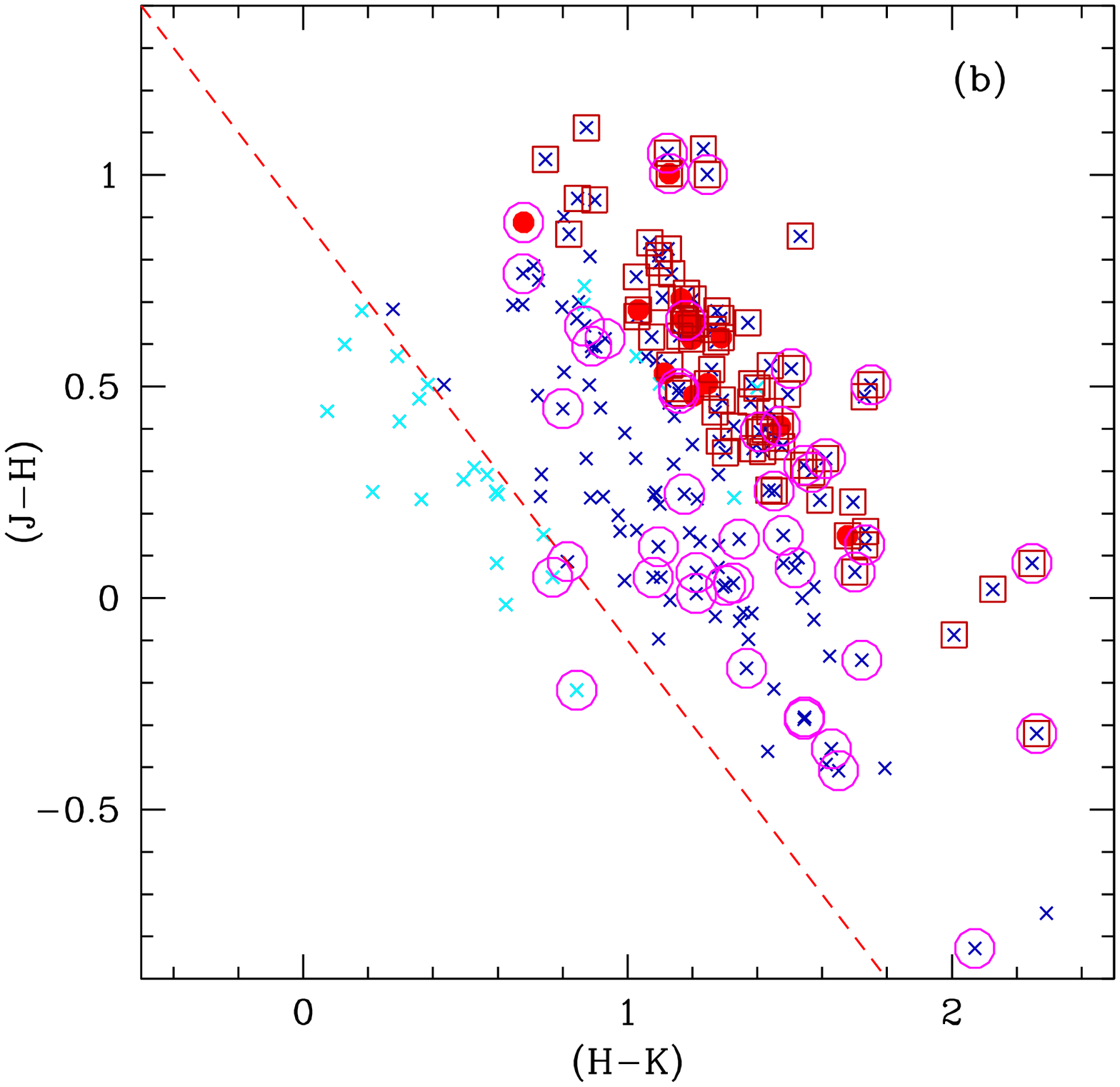}
\caption{Colour-magnitude relation (a) and colour-colour diagram (b) of cluster candidates and confirmed members. Left panel: $(J-K)$ vs. $K$ magnitude; Right panel: $(J-K)$ vs. $(H-K)$ colour. Symbols are as follows: Red solid circles - spectroscopically confirmed cluster members; cyan crosses - likely stars and foreground objects ($K<17.4$); blue crosses - all other detections; magenta circles - excess emitters; red boxes - red sequence galaxies. The cluster red sequence fit of \citet{Lidman2008} is shown as a long-dashed line. Red sequence galaxies are defined as galaxies redder than $0.2$ magnitudes blueward of this red sequence fit. The error bars in panel (a) indicate average photometric errors in each magnitude bin. \label{fig:color-color}}
\end{figure*}
%----------------------------- end Figure 3 ----------------------------------------

Figure~\ref{fig:color-color} shows all sources in the near-infrared colour-magnitude and colour-colour diagrams.  \citet{Lidman2008} identified a tight correlation between near-infrared red colour and brightness for galaxies in the XMMU~2335 cluster which is a result of galaxies harbouring old, evolved stellar populations.  They obtained a fit of the cluster's red sequence, which is plotted as long-dashed line in Figure~\ref{fig:color-color}.  Galaxies that are redder than $0.2$ magnitudes blueward of the red sequence are marked with red squares in Figure~\ref{fig:color-color}, and account for 45\% of our sample. These galaxies are likely to be cluster members since the likelihood to find foreground galaxies in this colour range is very low, as shown by \citet{Lidman2008}. This sample of galaxies on the red sequence is considered separately in the following analysis and is marked in Figures~\ref{fig:color-color}, \ref{fig:sfr-hmag} and \ref{fig:sfr-dist} with red squares.

Galaxies with excess line emission and hence star formation show a broad range of $(J-K)$ colours and are distributed over the whole colour-magnitude space. We also find objects in our sample with colours redder than the red sequence. In fact, a large fraction ($\sim 60\%$) of these extremely red objects are excess emitters (magenta circles in Figure~\ref{fig:color-color}), whereas only $\sim20 \%$ of all galaxies are classified as excess emitters. One possibility is that these galaxies are dust-reddened star-forming galaxies rather than red due to old stellar populations.

We also use the colour-magnitude and colour-colour diagrams to try to identify any interloping stars in the narrow-band sample. Star are distinguished from background galaxies by showing significantly bluer colours, typically below $(J-K) < 0.9$. This limit is shown in both panels in Figure~\ref{fig:color-color} as a short-dashed line. The left panel of Figure~\ref{fig:color-color} reveals a well-defined stellar locus at the bright end, whereas contamination from extremely blue (and hence fainter) galaxies is possible at the faint end. However, very few objects in our sample are close to this limit. The three excess emitting objects that are located below the colour limit of the stellar locus could be such very blue, star-forming galaxies. However, without spectroscopic information we cannot establish their cluster membership.

\subsection{Final galaxy sample selection}\label{finalsample}

The full sample of detected sources comprises 201 objects, however, in the analysis in the following sections of this article we exclude objects unlikely to be cluster members according to the following criteria. Likely stars identified via their colour ($(J-K) < 0.9$) described above, and via the source extraction parameter CLASS\_STAR, which uses the light distribution and width of the point spread function to identify stars. In order to minimise the chance of excluding unresolved galaxies, we require this parameter to be CLASS\_STAR $> 0.8$ in the $H$-band images and CLASS\_STAR $> 0.98$ in the higher resolution $K$-band images. Likely foreground galaxies are also excluded through their brightness, cross-checked by visual inspection of the images. We exclude all objects that are brighter than the brightest cluster galaxy in $K$-band ($K<17.4$), which affects eight objects in our sample. All of these are resolved galaxies with disk like morphologies and significantly larger sizes than the brightest cluster galaxy.  We therefore conclude that these objects are most likely foreground galaxies and remove them from the analysis. None of the objects excluded due to their brightness are excess emitters and all of their measured SFRs are consistent with zero within the errors.

After applying the above cuts, we obtain a final sample of 163 objects, which are listed in Table~\ref{tab1}, whereas all excluded objects are shown in Table~\ref{tab2} (the full tables are available in electronic form only). Note that the comparison of our data with the HAWK-I images revealed a small offset in the astrometric solution, which we have corrected by matching bright sources in the two images. This results in slightly different coordinates listed in Tables~\ref{tab1} and \ref{tab2} and the coordinates listed for the same objects in Table~1 of \citet{Bau11a}.

%------------------------------- begin Table 1 ----------------------------------------------------
\begin{table*}
\caption{Properties of detected objects in both pointings, cluster centre and outskirts. Spectroscopically confirmed members are given first, separated by a horizontal line. The full version of this table is available in electronic form only.\label{tab1}}
\begin{tiny}
\begin{tabular}{lllcccccc}
\hline
ID & $\alpha$ & $\delta$ & $H_{broad}$ & $H_{narrow}$ & SFR & (J-K) & dist. BCG & excess \\
   & [hh mm ss] & [dd mm ss] & {\tt MAG\_APER} & {\tt MAG\_APER} & [$M_\odot$yr$^{-1}$] & {\tt MAG\_APER} & [kpc] & $\Sigma$ \\
\hline

1	&	22	35	20.8	&	-25	57	40	&	20.48	$\pm$	0.03	&	20.57	$\pm$	0.06	&	0.0			&	1.83	$\pm$	0.02	&	0	&	-2.9	 \\
2	&	22	35	20.7	&	-25	57	38	&	20.80	$\pm$	0.04	&	20.86	$\pm$	0.06	&	0.0			&	1.86	$\pm$	0.03	&	24	&	-1.5	 \\
3	&	22	35	20.9	&	-25	57	36	&	21.41	$\pm$	0.04	&	21.56	$\pm$	0.07	&	0.0			&	1.81	$\pm$	0.04	&	35	&	-2.0	 \\
4	&	22	35	20.7	&	-25	57	44	&	20.91	$\pm$	0.04	&	21.11	$\pm$	0.06	&	0.0			&	1.87	$\pm$	0.03	&	44	&	-4.2	 \\
5	&	22	35	21.6	&	-25	57	26	&	21.74	$\pm$	0.05	&	21.64	$\pm$	0.07	&	0.8	$\pm$	0.7	&	1.65	$\pm$	0.04	&	159	&	1.1	 \\
6	&	22	35	19.0	&	-25	57	51	&	21.90	$\pm$	0.05	&	21.46	$\pm$	0.07	&	3.3	$\pm$	0.7	&	1.88	$\pm$	0.04	&	252	&	5.0	 \\
7	&	22	35	21.8	&	-25	57	14	&	22.50	$\pm$	0.07	&	21.44	$\pm$	0.07	&	6.2	$\pm$	0.7	&	2.13	$\pm$	0.10	&	257	&	9.5	 \\
8	&	22	35	22.8	&	-25	57	06	&	21.66	$\pm$	0.05	&	21.44	$\pm$	0.07	&	1.8	$\pm$	0.7	&	1.68	$\pm$	0.03	&	384	&	2.8    \\
9	&	22	35	22.0	&	-25	56	57	&	21.24	$\pm$	0.04	&	21.13	$\pm$	0.06	&	1.4	$\pm$	0.9	&	1.85	$\pm$	0.02	&	398	&	2.0	 \\
10	&	22	35	18.1	&	-25	58	06	&	21.84	$\pm$	0.05	&	21.39	$\pm$	0.07	&	3.5	$\pm$	0.7	&	1.57	$\pm$	0.08	&	422	&	5.4	 \\
11	&	22	35	25.5	&	-25	56	52	&	20.50	$\pm$	0.12	&	20.58	$\pm$	0.04	&	0.0			&	1.83	$\pm$	0.02	&	646	&	-2.6	 \\
12	&	22	35	22.8	&	-25	56	25	&	21.76	$\pm$	0.12	&	21.57	$\pm$	0.08	&	1.4	$\pm$	1.1	&	1.75	$\pm$	0.04	&	659	&	2.2    \\
13	&	22	35	27.1	&	-25	56	35	&	21.85	$\pm$	0.12	&	21.79	$\pm$	0.09	&	0.4	$\pm$	1.0	&	1.71	$\pm$	0.04	&	905	&	0.6	 \\
14	&	22	35	21.3	&	-25	55	21	&	21.46	$\pm$	0.12	&	21.20	$\pm$	0.06	&	2.6	$\pm$	1.3	&	1.91	$\pm$	0.05	&	1190	&	4.1$^{1}$ \\
\hline																														
15	&	22	35	25.4	&	-25	55	58	&	21.15	$\pm$	0.12	&	21.17	$\pm$	0.06	&	0.0			&	1.50	$\pm$	0.03	&	10	&	-0.4	 \\
16	&	22	35	28.0	&	-25	56	15	&	20.77	$\pm$	0.12	&	20.93	$\pm$	0.05	&	0.0			&	1.89	$\pm$	0.04	&	11	&	-3.9	 \\
17	&	22	35	22.2	&	-25	55	08	&	23.92	$\pm$	0.28	&	23.74	$\pm$	0.52	&	0.2	$\pm$	0.6	&	1.03	$\pm$	0.14	&	13	&	0.3	 \\
18	&	22	35	20.8	&	-25	57	42	&	23.37	$\pm$	0.12	&	23.32	$\pm$	0.23	&	0.1	$\pm$	0.4	&	1.55	$\pm$	0.08	&	20	&	0.1	 \\
19	&	22	35	21.1	&	-25	57	41	&	22.52	$\pm$	0.07	&	22.59	$\pm$	0.13	&	0.0			&	1.37	$\pm$	0.07	&	30	&	-0.4	 \\
20	&	22	35	20.8	&	-25	57	44	&	22.05	$\pm$	0.06	&	22.19	$\pm$	0.10	&	0.0			&	1.83	$\pm$	0.06	&	36	&	-1.1	 \\
\multicolumn{9}{l}{...}	\\																													
\hline
\end{tabular} \\
%\begin{flushleft}
$^{1}$ Not defined as excess emitters due to their low H$\alpha$ equivalent width ($EW < 20\mathrm{\AA}$, see Section~\ref{sec:contaminants})
%\end{flushleft}
\end{tiny}
\end{table*}
% ------------------------------------- end Table 1 ---------------------------------------

%------------------------------- begin Table 2 ----------------------------------------------------
\begin{table*}
\caption{Properties of objects excluded from the sample due to $(J-K)$ colour, $K$-band magnitude or CLASS\_STAR parameter (see text). The full version of this table is available in electronic form only.\label{tab2}}
\begin{tiny}
\begin{tabular}{lllccccc}
\hline
ID & $\alpha$ & $\delta$ & $H_{broad}$ & $H_{narrow}$ & (J-K) & CLASS\_STAR & CLASS\_STAR \\
   & [hh mm ss] & [dd mm ss] & {\tt MAG\_APER} & {\tt MAG\_APER} & {\tt MAG\_APER} & H & K  \\
\hline
26	&	22	35	21.6	&	-25	57	49	&	23.76	$\pm$	0.16	&	24.45	$\pm$	0.63	&	0.73	$\pm$	0.15	&		0.36	&	0.35	\\
53	&	22	35	22.1	&	-25	57	14	&	23.24	$\pm$	0.11	&	23.32	$\pm$	0.23	&		0.85	$\pm$	0.08	&		0.36	&	0.41	\\
62	&	22	35	20.2	&	-25	58	17	&	22.99	$\pm$	0.09	&	22.99	$\pm$	0.17	&		0.61	$\pm$	0.08	&		0.42	&	0.48	\\
63	&	22	35	21.3	&	-25	58	17	&	21.61	$\pm$	0.05	&	21.38	$\pm$	0.07	&		0.77	$\pm$	0.02	&		0.73	&	0.84	\\
66	&	22	35	21.0	&	-25	57	01	&	23.51	$\pm$	0.13	&	23.93	$\pm$	0.39	&		0.68	$\pm$	0.09	&		0.42	&	0.46	\\
%68	&	22	35	22.0	&	-25	58	16	&	24.68	$\pm$	0.33	&	23.18	$\pm$	0.20	&		 99	$\pm$	99	&		0.36	&	99	\\
%75	&	22	35	19.6	&	-25	58	22	&	22.97	$\pm$	0.09	&	21.92	$\pm$	0.08	&		 99	$\pm$	99	&		0.49	&	99	\\
%85	&	22	35	17.9	&	-25	57	05	&	22.20	$\pm$	0.06	&	21.96	$\pm$	0.08	&		0.60	$\pm$	0.05	&		0.51	&	0.59	\\
%91	&	22	35	21.3	&	-25	56	34	&	23.83	$\pm$	0.26	&	25.96	$\pm$	4.03	&		 99	$\pm$	99	&		0.35	&	99	\\
%94	&	22	35	20.8	&	-25	56	31	&	18.76	$\pm$	0.12	&	18.69	$\pm$	0.03	&		0.86	$\pm$	0.00	&		0.97	&	0.98	\\
\multicolumn{8}{l}{...} \\
\hline
\end{tabular}
\end{tiny}
\end{table*}
% ------------------------------------- end Table 2 ---------------------------------------

Out of the final sample of 163 objects 14 are spectroscopically confirmed cluster members (listed first in Table~\ref{tab1}) and 32 are narrow-band excess emitters and therefore the most likely cluster members, although the interloper fraction could be as high as $\sim 30\%$ according to the estimate of \citet{Sobral11}. We further constrain the sample of likely cluster members through near-infrared colours using the fit to the red sequence of the cluster colour-magnitude relation by \citet{Lidman2008}. Red sequence galaxies are defined as galaxies with (J-K) colours redder than 0.2 magnitudes blueward of the \citet{Lidman2008} fit and are likely to be cluster members, as described in Section~\ref{sec:redsequence}. We find that 75 galaxies (or 46\%) of our sample are classified as red sequence galaxies. This sub-sample of galaxies is also considered separately in the following analysis.

We also note that a significant fraction of the non-emitters in our sample will not be cluster members. However, considering the large fraction of galaxies located on the cluster red sequence and the fact that most spectroscopically confirmed cluster members do not show excess emission (11 out of 14, or $\sim$80\%) we can infer that at least some of the non-emitters will indeed be members of the cluster. In order to avoid biasing our sample towards star-forming galaxies we include all the 163 galaxies in the analysis throughout the rest of this article, but we consider excess emitters as well as galaxies close to the cluster red sequence separately, since they have a higher likelihood to be cluster members.

%%%%%%%%%%%%%%%%%%%%%%%%%%%%%%%%%%%%%%%%%%%%
%%                   Results              %%
%%%%%%%%%%%%%%%%%%%%%%%%%%%%%%%%%%%%%%%%%%%%
\section{Results}\label{sec:results}

In the following we present the star formation rates (SFRs) derived from the $H$ narrow-band imaging and investigate any possible correlations between the star-forming properties of the galaxies and SFR and their magnitude, distance from the cluster centre (as defined by the BCG), and stellar mass. This work extends the analysis presented in \citet{Bau11a} to much larger cluster-centric radii (1.5 Mpc here compared to the 500 kpc in \citet{Bau11a}) as well as over a larger galaxy sample. Additionally we make use of HAWK-I imaging of the cluster in the $J$ and $K$ bands described in \citet{Lidman2008}. 

Table~\ref{tab1} summarises the resulting SFRs, magnitudes, colours and distances from the cluster centre for all objects we consider in this study, with spectroscopically confirmed cluster members listed first. The last column in Table~\ref{tab1} identifies excess line-emitters via the $\Sigma$-level of excess emission as described in Section~\ref{sec:contaminants}. The $\Sigma$-level is computed from the observed limiting magnitudes given in Section~\ref{sec:detection} and is set to zero if the required minimum equivalent width of 20$\mathrm{\AA}$ is not reached.

\begin{figure}
\includegraphics[width=0.465\textwidth]{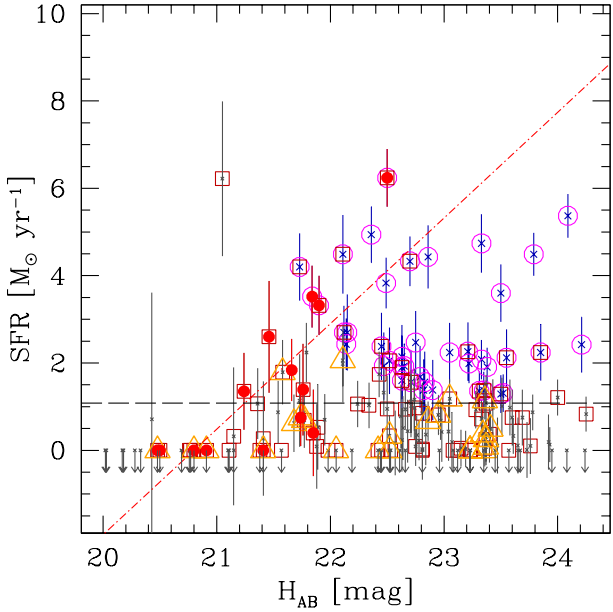}
\caption{Star formation rates as a function of $H$ broad-band magnitude. The symbols are as follows: red solid circles - spectroscopically confirmed member galaxies; magenta circles - excess line-emitters; red boxes - galaxies on the red sequence; grey crosses - all other detections; orange triangles - galaxies within 200~kpc of the BCG. Arrows represent galaxies with fainter narrow-band magnitudes than expected from their broad band magnitude, a possible evidence for H$\alpha$ in absorption. The red dashed-dotted line represents a least squares fit to the data points of confirmed members. The $5\sigma$ SFR detection limit is shown as the dashed horizontal line, as described in Section~\ref{sec:detection}. \label{fig:sfr-hmag}}
\end{figure}

Section \ref{sec:sfr-mag} presents the derived SFRs as a function of magnitude (Figure~\ref{fig:sfr-hmag}) and cluster-centric distance (Figure~\ref{fig:sfr-dist}), analogous to Figures 4 and 5 in \citet{Bau11a}.  We include the data of the cluster centre in this analysis and present it together with the new observations of the cluster outskirts.  In Section \ref{SSFR} we estimate the specific star-formation rates (SSFR), i.e., the star-formation rate per unit of stellar mass, using a mass-to-light ratio applied to the deep $K$-band observations. 

%--------------------------- begin Figure 4: SFR-Hmag ------------------------------------------
%----------------------------- end Figure 4 ----------------------------------------

\subsection{SFR as a function of magnitude and cluster-centric distance}\label{sec:sfr-mag}

We first analyse the correlation between SFR and an intrinsic galaxy property, galaxy magnitude. Figure~\ref{fig:sfr-hmag} shows individual SFRs as a function of $H$ broad-band magnitude, which corresponds to a rest-frame wavelength of $\sim$6900$\mathrm{\AA}$.  The different subsamples defined in Section~\ref{finalsample} are plotted separately: confirmed cluster members are shown as red dots, while all other objects are plotted as crosses. Additional symbols denote galaxies defined as excess emitters (blue with open magenta circles), galaxies on the red sequence (open red squares) and galaxies within the central $R_{_Q}=200$ kpc (orange triangles). Galaxies with negative $(H-H_{narrow})$ colours are marked with grey arrows.

We confirm the correlation between SFR and $H$-band magnitude for confirmed cluster members found in the cluster centre in our previous study \citep{Bau11a}. A linear fit to the data of confirmed cluster members is plotted as dashed-dotted line. Brighter galaxies tend to have less (or no) star formation. The statistical significance of this correlation is 2$\sigma$, given by the Spearman rank correlation coefficient ($\rho=0.85$) and its corresponding significance level for our sample size. The probability that $H$ band magnitude and SFR are uncorrelated is then $P=4\%$.  Assuming that the amount of star formation per unit stellar mass in a galaxy is constant - and that galaxy light traces stellar mass - we would expect that brighter galaxies exhibit higher SFRs. Our data however suggests the opposite: brighter (more massive) galaxies have less star-formation and older stellar populations, while fainter (less massive) galaxies are still forming. This trend is commonly referred to as downsizing \citep[e.g.][]{Cowie96} and will be further discussed in Section~\ref{SSFR}.

%--------------------------- begin Figure 5: SFR-distance ------------------------------------------
\begin{figure}
\includegraphics[width=0.465\textwidth]{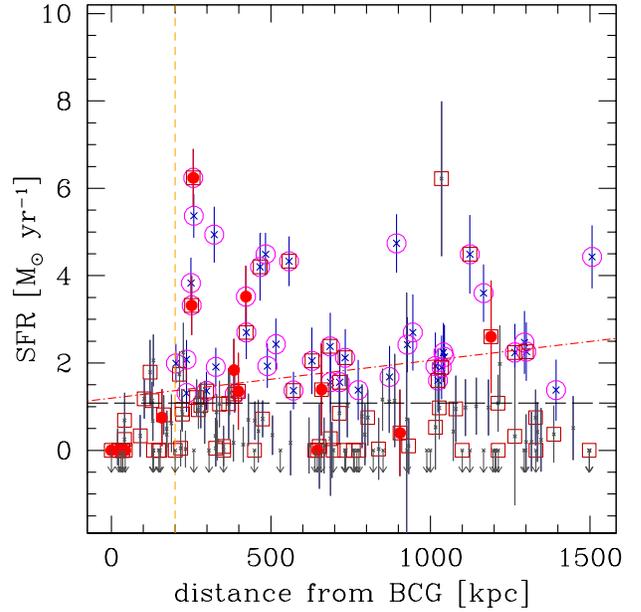}
\caption{Star formation rates as a function of distance from the cluster centre. Symbols are the same as shown in Figure~\ref{fig:sfr-hmag}. The vertical dashed line marks the projected distance within which no star formation occurs ($R_{_Q}$). Galaxies within this limit are also marked with orange triangles in Figure~\ref{fig:sfr-hmag}.  
\label{fig:sfr-dist}}
\end{figure}
%----------------------------- end Figure 5 ----------------------------------------

We find no correlation between SFR and brightness for the excess emitters (magenta circles, $P=35\%$) or for galaxies on the red sequence (red squares, $P=65\%$). Star formation activity has already ceased in the brightest galaxies in the cluster, which are located on the bright end of the red sequence, whereas objects that are still forming stars do so independently of their brightness. It is worth noting however that the SFRs are overall low, much lower than expected for galaxies at the cluster's redshift \citep[see e.g.][]{Bau11b}, which will be discussed in more detail in the following section.  We find a median SFR for all galaxies in our sample of $<$SFR$>$ = 0.7$~\pm~0.8$~M$_\odot$yr$^{-1}$, although a significant amount of interlopers could artificially decrease this value.  If we consider only galaxies located around the cluster red-sequence (see Section~\ref{sec:redsequence}), which are more likely to be cluster members than the general sample, we obtain a median $<$SFR$>$ = 0.8$~\pm~0.8$~M$_\odot$yr$^{-1}$, which is very similar to the value for the full sample. Taking only spectroscopically confirmed cluster members into account, we measure a median of $<$SFR$>$ = 1.3$~\pm~1.9$~M$_\odot$yr$^{-1}$, which is still low compared to the average SFR of field galaxies at similar redshifts.

To further compare the star-formation activity within the cluster with the general field environment at the same redshift we compute the SFR density in 
M$_\odot$yr$^{-1}$ per cubic Mpc. Due to the limited coverage of our observations in the cluster outskirts we restrict this analysis to the cluster centre. We assume that the observed volume can be approximated by a cylinder with the radius covered by the central pointing ($\sim$ 0.5 Mpc) and a depth of 1.5 Mpc, corresponding to $\sim$1.5 times the virial radius of the cluster \citep{Rosati2009}. The total SFR is then the sum of the individual SFRs of all galaxies detected within this area, resulting in SFR densities of $\sim$50~M$_\odot$yr$^{-1}$Mpc$^{-3}$ for excess emitters (based on 19 galaxies) and $\sim$15~M$_\odot$yr$^{-1}$Mpc$^{-3}$  for confirmed cluster members (based on 10 galaxies). This is a factor of $\sim$10 higher than the SFR density of the universe at $z\sim1.4$ of $\sim$0.2~M$_\odot$yr$^{-1}$Mpc$^{-3}$ \citep{HopBea06} due to the high galaxy number density of the cluster. Note however that the uncertainty of this estimate is very high due to the uncertainty of the individual SFRs ($\sim 30\%$), the cluster volume as well as the contamination rate of up to $30\%$ in the sample of excess emitters.

Our previous study of galaxies in the cluster centre \citep{Bau11a} revealed a lack of star-formation within the inner 200 kpc of the cluster, and identified this as a quenching radius, $R_{_Q}$.   We also noted the possibility for a steady increase in SFR with cluster-centric radius.  With the new observations shown in Figure~\ref{fig:sfr-dist} we can now trace activity three times farther out to a radius of 1.5 Mpc. However, we do not see evidence for an increase in SFR with distance from the cluster centre.  Using again the Spearman rank correlation test we find that the probability that cluster-centric distance and SFR are uncorrelated is $P=66\%$ for confirmed members, $P=36\%$ for excess emitters and $P=42\%$ for all candidates. We find that outside $R_{_Q}$, the SFR of all galaxies with evidence of star formation remain at a constant level of $<$SFR$> =$ 2.7~$\pm~1.0$~M$_\odot$yr$^{-1}$.  This is consistent with the average SFR found by \citet{Hayashi10} for cluster galaxies in the  XMMXCS J2215.9-1738 cluster at $z= 1.46$. Note also that the average SFR value of cluster galaxies is much lower than the average value found for field galaxies at this redshift \citep{Daddi10,Bau11b}, as discussed further in the following section.

\subsection{Specific star formation rates}\label{SSFR}

In the following we use the observed $K$-band magnitudes to obtain an estimate of the galaxy stellar mass.
The central wavelength of the observed $K$-band represents a rest-frame wavelength of roughly $9200\mathrm{\AA}$, representative of the underlying evolved stellar population, and less affected by dust extinction than the shorter wavelength $H$-band whose rest-frame central wavelength is $\sim 6900\mathrm{\AA}$. We calculate a stellar mass-to-light ratio ($M/L$) by assuming an evolved stellar population of the age of the Universe at the cluster's redshift (4.5 Gyr) combined with a 5\% by mass contribution from a young stellar population using the models of \citet{Worthey94}. This relatively small contribution of young stars is motivated by the low star formation rates of the galaxies in our sample. 

We use exponentially declining star formation histories of the form SFR~$=$~SFR$_0~e^{(-t/\tau)}$, with a star formation timescale of $\tau = 1 \times 10^9$ yr.  Integrating over the last billion years, we find that an average galaxy with active star formation typical of galaxies within this sample ($\sim 2 M_\odot$yr$^{-1}$) would have created $\sim3 \times 10^9 M_\odot$ of stars over the last 10$^9$ years.  Note that the $M/L$ of a 4.5 Gyr old population at the rest-frame wavelength of $9200\mathrm{\AA}$ is not significantly affected by changing the amount of young stars up to 10\% \citep{Worthey94}. It changes from $M/L\sim1.4$ assuming a 10\% contribution of young stars to $M/L\sim1.5$ for a contribution of 1\%.
We adopt a mass-to-light ratio of $M/L=1.5$ to obtain an estimate for the galaxy stellar mass (M$^\ast$) from the observed K-band magnitude.  From this we estimate SSFRs as SSFR~=~SFR/M$^\ast$.

%--------------------------- begin Figure 6: SSFR ------------------------------------------
\begin{figure}
\includegraphics[width=0.465\textwidth]{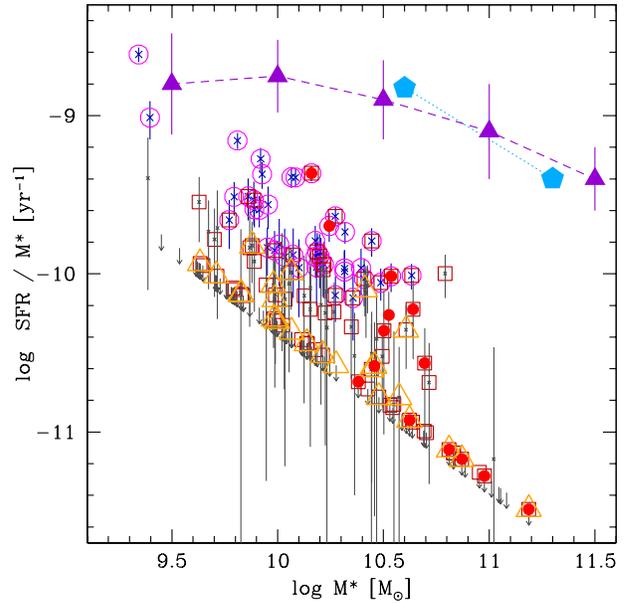}
\caption{Specific star formation rate as a function of stellar mass. Symbols are the same as shown in Figure~\ref{fig:sfr-hmag}. Upper limits for galaxies without detected star-formation (SFR$~<0.5$M$_\odot$yr$^{-1}$) are shown as grey arrows. The average SSFRs of a general field sample at similar redshift is plotted as blue pentagons and dark purple triangles \citep{PG2008,Bau11b}. \label{fig:ssfr}}
\end{figure}
%----------------------------- end Figure 6 ----------------------------------------

The resulting specific star formation rates are shown in Figure~\ref{fig:ssfr} as a function of stellar mass. 
The symbols are the same as in Figure~\ref{fig:sfr-hmag}, with confirmed members plotted as red dots, all other objects as crosses, excess emitters in blue with magenta circles, galaxies on the red sequence with red squares and galaxies within the central $R_{_Q}=200$ kpc marked with orange triangles. Upper limits are plotted as arrows for galaxies with SFR~$< 0.5$ M$_\odot$yr$^{-1}$. 

Figure~\ref{fig:ssfr} shows that among galaxies in the XMMU2335 cluster, higher stellar mass galaxies have lower SSFRs, a common result from broader field galaxy samples at lower redshift \citep{Cowie96,BE00,Bauer05,Juneau2005}.  Regardless of their position in the cluster, star formation is effectively shut off for all massive galaxies with log M$^\ast$/M$_\odot$~$>10.75$.  Galaxies within the central $R_{_Q}=200$ kpc have a wide range of stellar masses, but show a clear offset towards low (below average) or only limiting values of SSFR.  For galaxies with log M$^\ast$/M$_\odot$~$<10.75~$, the tendency is for those within $R_{_Q}=200$ kpc to have lower SSFRs. A K-S test gives a probability of $P=5\%$ that the SSFRs of galaxies inside and outside of $R_{_Q}$ follow the same distribution.
Galaxies with lower stellar mass and outside of the quenching radius are building up stellar mass via star formation, while massive galaxies or those in the cluster core are not. 
 
In Figure~\ref{fig:ssfr} we also show average SSFRs of typical field galaxies at this redshift ($1.5 < z < 2$). These samples are drawn from \citet{Bau11b} (triangles) and \citet{PG2008} (pentagons), the former being based on SFRs determined from rest-frame ultraviolet (UV) observations, while the latter uses SFRs determined from infrared (IR) data. The rest-frame UV star-formation rates are based on Hubble Space Telescope observations in $z$-band and are significantly deeper than the IR observations, reaching a limiting SFR of $~0.3~$M$_\odot$yr$^{-1}$ at $z=1.5$, comparable to the limiting SFR in this study. Although the scatter between the individual SFRs derived from UV and IR data are considerable \citep{Bau11b}, the average SFRs (and SSFRs) of the whole sample are consistent.  Figure~\ref{fig:ssfr} shows that the median values of both, UV and IR based SSFRs of the  \citet{Bau11b} study are clearly offset from the SSFRs of galaxies in our cluster sample by a factor of $\sim$5-10.

We conclude that galaxies in the cluster centre have lower SSFRs than the rest of the cluster galaxies, which in turn have lower SSFRs than field galaxies at the same redshift by a factor of a few for log M$^\ast$/M$_\odot$~$<10.75~$ galaxies, and up to a factor of $>10$ for massive galaxies.

%%%%%%%%%%%%%%%%%%%%%%%%%%%%%%%%%%%%%%%%%%%%
%%                   Discussion              %%
%%%%%%%%%%%%%%%%%%%%%%%%%%%%%%%%%%%%%%%%%%%%
\section{Discussion}\label{sec:discussion}

We have presented a detailed look at the properties of individual galaxies in the massive galaxy cluster, XMMU2235.3-2557, at $z=1.39$.  The number of studies of individual galaxy clusters at high redshifts is increasing and as such, it is important to compare results presented here to other distant clusters to see if consistent results are being found.  

It remains challenging to solidly identify clusters at $z>2$, but several studies have found high concentrations of sources around radio galaxies \citep{Pentericci1997,Kurk2004,Venemans2007,Tanaka2011,Hatch2011} or from over-densities in infrared observations \citep{Gobat2011}.  Although it is not likely that these are virialized systems, they show a variety of star formation activity in their cores, from galaxies with strong emission lines \citep[e.g.][]{Tanaka2011} to those that show red galaxies with old stellar populations \citep[e.g.][]{Gobat2011}.  It is evident that many of these clusters are still in an active phase of the galaxy formation process and it is most likely that these types of systems will become something like XMMU2235 at $z=1.39$ and eventually the most massive clusters observed in the local universe.   
 
In recent years many clusters in the redshift range of $1.3<z<1.75$ have been found and studied in detail.  Again, a variety of techniques have been used to detect these massive high redshift systems, but interestingly, whether or not the galaxies in their cores harbour exclusively old stellar populations with no evidence of new star formation, seems to be dependent on how the cluster was detected.  For instance, almost all clusters detected so far by X-ray emission show evidence for a lack of star formation activity in their cores based on narrow-band observations \citep{Bau11a}, or by using photometric colours to identify red galaxy populations coincident with the peak of the X-ray emission \citep{Henry2010,Hilton2010,Fassbender2011,Nastasi2011,Santos2011}. This is probably due to the correlation between X-ray luminosity and total cluster mass and the fact that the most massive structures are the first to form. The more massive a structure is the earlier it collapses, giving the environment more time to transform galaxies via ram pressure stripping \citep{gunn-gott1972} or merging \citep[e.g.][]{Mihos94}.  Similarly, the higher the density of the hot intra cluster medium (ICM), the more effective is the stripping of the cold gas within galaxies and hence the quenching of star formation \citep[see also][]{Larson80}.  

A notable exception to this statement is the study of \citet{Hayashi10} of the X-ray detected cluster XMMXCS J2215.9$-$1738 at $z = 1.46$. The authors find no decrease in the emission of narrow-band [OII] within galaxies in the cluster core and interpret this as evidence of AGN activity that suppresses star formation among galaxies in the cluster.  In a different study of the same cluster (XMMXCS J2215.9$-$1738), \citet{Hilton2010} see evidence of star formation in galaxies at the exact cluster-centric radius we have identified as $R_{_Q}~\sim~200$~kpc, but no star formation within that radius. Also, in a recent study of the X-ray detected galaxy cluster XMMU J1007.4$+$123 at $z=1.55$ \citet{Fassbender2011} conclude that although the red sequence is not yet in place in the core of the cluster, there is a spatial density peak of red galaxies coincident with the X-ray emission.

Clusters detected from methods other than X-ray emission, likely to be non-virialized, less massive structures, seem to show no evidence for having exclusively red or non-star-forming galaxies at their core.  For example, the cluster ClG J0218.3$-$05101 at $z=1.62$ was discovered from {\it Spitzer} infrared observations by \citet{Papovich2010} and a follow-up study by \citet{Tran2010} finds a high level of star formation activity in the core of this cluster based on {\it Spitzer}-based IR SFRs and SED-fitting. Recently \citet{Pierre11} investigated the X-ray emission of this cluster and do not find strong evidence for extended X-ray emission. Another instance comes from an over-density in the GMASS redshift distribution found by \citet{Kurk2009}. They identify a massive structure called Cl 0332-2742 at $z=1.6$, and while the galaxies in this structure show a strong colour bimodality, it is not conclusive that the red galaxies concentrate towards the centre of the region.  

We conclude that our result of suppressed star-formation across the entire cluster but specifically within the central region of XMMU2335 is in agreement with the results found for most other massive X-ray detected clusters at high redshift. However, for the first time we can demonstrate with the measurements of individual SFRs that the suppression of star-formation is an effect of the cluster environment which persists at fixed galaxy stellar mass.
One issue to consider when interpreting these results is that our observations of the outskirts of XMMU2335 from R$=800 - 1500$~kpc only cover one diagonal region outward from the cluster core and so do not encompass the entire annulus.  This region was targeted because it hosts a set of  spectroscopically confirmed cluster members.  It is possible that the presence of a filament or lack of structure in this region might bias our results in some way. It is more likely, however, that we miss star-forming galaxies in the cluster outskirts, a fact that would not change the above conclusions.

%%%%%%%%%%%%%%%%%%%%%%%%%%%%%%%%%%%%%%%%%%%%%%%%%%%%%%%
%%                   Summary and Conclusions         %%
%%%%%%%%%%%%%%%%%%%%%%%%%%%%%%%%%%%%%%%%%%%%%%%%%%%%%%%
\section{Summary and conclusions}\label{sec:summ}

We determine star formation rates (SFRs) in the most massive galaxy cluster at high redshift ($z>1$) known to date,  XMMU2235.3-2557 at $z=1.39$ \citep{Mullis05,Rosati2009}. This cluster was previously found to be mainly composed of an evolved galaxy population through the analysis of broad band colours \citep{Lidman2008} or spectral stacking techniques \citep{Rosati2009}. This is the first study to determine the star formation activity of individual galaxies out to large cluster radius of 1.5~Mpc.  We use deep narrow-band imaging obtained using NIRI on the Gemini North Telescope, which corresponds to the wavelength of the H$\alpha$ emission line at the redshift of the cluster. We find the following results:

\begin{enumerate}
\renewcommand{\theenumi}{\arabic{enumi}.}

\item  We directly measure the SFRs of individual objects out to a cluster radius of $\sim 1.5$ Mpc down to a sensitivity of SFR$\sim1$M$_\odot$yr$^{-1}$.  Excluding likely stars and foreground galaxies we find a sample of 163 galaxies over a surface area of $\sim$7.45 arcmin$^{2}$ ($\sim$ 1.7 Mpc$^2$), of which 14 galaxies are spectroscopically confirmed cluster members. A total of 32 out of the 163 galaxies are identified as excess line-emission galaxies and therefore the most likely to be cluster members. We also find a high fraction of galaxies in our sample (46\%) consistent with being on the cluster's well developed red sequence, which are likely to be passive cluster members with evolved stellar populations.\\

\item  Including all observed galaxies, we measure a median SFR of 0.7~M$_\odot$yr$^{-1}$ with a scatter of 0.8, consistent with the SFRs found for intermediate redshift clusters (e.g. EDisCS clusters at $0.4<z<0.8$) and a factor of $\sim$10 lower than field galaxies at $1<z<2$. Despite of the high fraction of contaminants in our sample, this statement is still valid if we consider only galaxies on the cluster's red sequence, which have a median SFR of $<$SFR$>$ = 0.8$~\pm~0.8$~M$_\odot$yr$^{-1}$, or spectroscopically confirmed members only ($<$SFR$>$ = 1.3$~\pm~1.9$~M$_\odot$yr$^{-1}$). Galaxies identified as excess emitters have a median star formation rate of $<$SFR$>$ = 2.7$~\pm~1.0$~M$_\odot$yr$^{-1}$. Note also that the highest SFRs in the cluster of $\sim$~6~M$_\odot$yr$^{-1}$ are still lower than the median SFR of field galaxies at this redshift, however it is worth noting that the dust correction remains a major source of uncertainty in this result.\\  

\item In the cluster core, all galaxies show suppressed or no evidence of star formation, regardless of stellar mass.  The environmental influence in the sample is that at fixed stellar mass, galaxies in the cluster core have lower SSFRs than the rest of the cluster galaxies, which in turn have lower SSFRs than field galaxies at the same redshift. We observe this effect out to $\sim 1.5$ Mpc, i.e. more than the virial radius of the cluster.\\

\item We do not see a continuous increase of SFR with distance from the cluster core. We argue that the star formation might not only be continuously suppressed as galaxies approach the cluster centre, but that there is a quenching radius ($R_{_Q}~\sim~200$~kpc) within which star formation is rapidly shut-off. This could be caused either by interaction with the ICM (ram pressure stripping) or frequent high velocity encounters (harassment) caused by the high galaxy density in the cluster centre where the crossing time at $R_{_Q}$ is 0.25 Gyr.\\

\item We find evidence that the suppression of star formation in the cores of early galaxy clusters might be related to how the cluster was detected. In particular the presence of extended X-ray emission from a hot ICM seems to suppress the star formation activity of cluster galaxies, whereas clusters that are detected via different methods harbour galaxies with active star formation throughout the cluster. This effect could be caused by processes related to the total mass of the parent structure as well as by the direct influence of the ICM on the gas content of cluster members.

\end{enumerate}

We conclude that star formation in this cluster is effectively shut off in the cluster centre already at $z=1.39$, when the universe was only $\sim$4.5 Gyr old. Galaxies at larger radii from the cluster centre are moderately forming stars, but not reaching the average SFR of field galaxies at this redshift. This suggests that the red sequence is built up from the inside out, starting at redshifts $z > 2$ in the most massive galaxy clusters.

\section*{Acknowledgments}

Based on observations obtained at the Gemini Observatory, which is operated by the Association of Universities for Research in Astronomy, Inc., under a cooperative agreement with the NSF on behalf of the Gemini partnership: the National Science Foundation (United States), the Science and Technology Facilities Council (United Kingdom), the National Research Council (Canada), CONICYT (Chile), the Australian Research Council (Australia), Minist\'erio da Ci\^encia e Tecnologia (Brazil) and Ministerio de Ciencia, Tecnolog\'ia e Innovaci\'on Productiva (Argentina).  The data presented in this paper originate from the Gemini programs GN-2007B-Q-79 and GN-2010B-Q-75, both observed in queue mode.  We would like to thank Chris Lidman for kindly providing the VLT images of the cluster which greatly improved the analysis presented here. We also thank the anonymous referee for providing very useful suggestions.

\bibliographystyle{mn2e}
\bibliography{halpha_out} 

\bsp

\label{lastpage}
\end{document}